\documentclass[a4paper,fleqn]{article} 
\usepackage{modsim}
\usepackage{times}
\usepackage{natbib} 
\usepackage{amsmath, amssymb, amsthm} 

\MODSIMhead{X. Luo and C. Foss, Inverse of magnetic dipole field using a reversible jump Markov chain Monte Carlo} 

\usepackage{rotating}
\usepackage{amsbsy,enumerate}
\usepackage{graphicx}
\usepackage{ccaption}

\makeatletter
\renewcommand{\fnum@figure}[1]{\textbf{\figurename~\thefigure}. }
\renewcommand{\fnum@table}[1]{\textbf{\tablename~\thetable}. }
\makeatother

\begin{document}

\title{Inverse of magnetic dipole field using a reversible jump Markov chain Monte Carlo}

\author{\underline{X. Luo} \address[A1]{\it{CSIRO Computational Informatics, Sydney, NSW Australia}},
{C. Foss} \address[A2]{\it CSIRO Earth Science and Resource Engineering, Sydney, NSW Australia}} 

\email{Xiaolin.Luo@csiro.au} 

\date{March 2009}

\begin{keyword}
 Magnetic dipoles, Markov chain Monte Carlo, reversible jump,
 trans-dimensional,  inverse problem
\end{keyword}

\begin{abstract}
    We consider a three-dimensional magnetic field produced by an arbitrary collection of dipoles.
    Assuming the magnetic vector or its gradient tensor field is measured above the earth surface,
    the inverse problem is to use the measurement data to find the location, strength, orientation and
    distribution of the dipoles underneath the surface. We propose a reversible jump Markov chain Monte
    Carlo (RJ-MCMC) algorithm for both the magnetic vector and its gradient tensor to deal with this
    trans-dimensional inverse problem where the number of unknowns is one of the unknowns.
    A special birth-death move strategy is designed to obtain a reasonable rate of acceptance for
    the RJ-MCMC sampling.

    Typically, a birth-move generates an extra dipole in the field.  In order to have a reasonable acceptance rate
for the birth move, we try to keep the change in the likelihood
function due to the extra dipole to be small. To achieve this small
perturbation in likelihood function, instead of randomly adding a
new dipole to the system, we replace one of the existing dipoles
with two new dipoles. Ideally, the combined magnetic field produced
by the two new dipoles should be very close to the magnetic field of
the replaced dipole, at every measurement point. It is analytically
difficult to ensure this closeness of magnetic field at every
measurement point.

We can simplify the problem by ensure that the magnetic field
produced by the new pair of dipoles is close to that of the old
dipole at one key measurement point, for example at the centre of
the measurement range. Typically the measurement points can be
arranged in a horizontal rectangular lattice and that key point can
be chosen to be located at the centre of the lattice. We  show
 that for any randomly chosen dipole to be removed, we can place two dipoles with the same strength at a special location
  such that the magnetic field at the key point remain exactly the same as before this two-for-one
replacement of the birth move. The two new dipoles are then
separated by random moves similar to that of a within-model move.
The death move is simply the reverse of the birth move.

Some preliminary results show the strength and challenges of the
algorithm in inverting the magnetic measurement data through
dipoles. Starting with an arbitrary single dipole, the algorithm
automatically produces a
    cloud of dipoles to reproduce the observed magnetic field, and
    the true dipole distribution for a bulky object is better
    predicted than for a thin object. Multi-objects located at
    different depths remain a very challenging inverse problem.

\end{abstract} 

\maketitle

\section{INTRODUCTION}

Monte Carlo techniques for geophysical inversion were first used
about forty years ago,  \citet{Keilis1967},
\citet{AnderssenAndSeneta1971},  \citet{Anderssen1972}. since then
there has been considerable advances in both computer technology and
mathematical methodology, and therefore an increasing interest in
those methods. Some examples can be found in
\citet{MosegaardTarantola2002}, \citet{MalinvernoLeaney2005},
\citet{SambridgeEtal2006}, \citet{BodinSambridge2009} and
\citet{Luo2010}.

There is a class of problems where the ``number of unknowns is one
of the unknowns". For these problems, a number of frameworks have
been developed since the mid-1990s to extend the fixed-dimension
Markov chain Monte Carlo (MCMC) to encompass trans-dimensional
stochastic simulation. Among these trans-dimensional schemes, the
reversible jump Markov chain sampling algorithm proposed by
\citet{Green1995} is certainly the most well understood and well
developed. A survey of the state of the art on trans-dimensional
Markov chain Monte Carlo can be found in \citet{Green2003}.
Trans-dimensional MCMC has been successfully applied to geophysical
models, see \citet{SambridgeEtal2006} and
\citet{BodinSambridge2009}. \citet{Luo2010} proposed a RJ-MCMC
algorithm to detect the shape of a geophysical object underneath the
earth surface from gravity anomaly data, assuming a two-dimensional
polygonal model for the object.

Although the idea of \citet{Luo2010} can in principle be extended to
three-dimensional cases with polygons replaced by polyhedrons, in
practice much numerical difficulties could be encountered. What is
more, an arbitrary three-dimensional real object cannot always be
presented by a simple polyhedron. Another limitation of the
development in \citet{Luo2010} is that it is not trivial to extend
the model to multiple objects. The present paper is the first
attempt to invert a three-dimensional magnetic dipole field using
RJ-MCMC.

\section{MAGNETIC FIELD AND LIKELIHOOD FUNCTION}

Consider an arbitrary magnetic dipole ${\rm {\bf m}}$ with magnitude
$m$ and unit vector ${\rm {\bf \hat {m}}}$, located at ${\rm {\bf
x}}(x,y,z)$, the magnetic field at an arbitrary point ${\rm {\bf
\tilde {x}}}(\tilde {x},\tilde {y},\tilde {z})$ is given by
\begin{equation}
{\bf H}({\bf m}, {\bf r})=-\mu_0 \nabla V({\bf m}, {\bf r}) = -
\frac{\mu_0m}{4\pi} \nabla \left ( \frac{\bf \hat m \cdot r}{ r^3}
\right ) = \frac{\mu_0m}{4\pi r^3} \left ( (3{\bf \hat m \cdot r} )
{\bf \hat   r} - {\bf \hat m}\right )
\end{equation}
\noindent where ${\rm {\bf r}} = {\rm {\bf \tilde {x}}}(\tilde
{x},\tilde {y},\tilde {z}) - {\rm {\bf x}}(x,y,z)$, $\mu _0 $ is the
magnetic permeability of free space.

Consider $k$ dipoles, each denoted as ${\rm {\bf m}}_i $, $i =
1,...,k$, and located at ${\rm {\bf x}}_i = (x_i ,y_i, z_i )$ and
with a strength $m_i $ and a direction unit vector ${\rm {\bf \hat
{m}}}_i $. Assume $N$ measurement locations at ${\rm {\bf \tilde
{x}}}_n = (\tilde {x}_n ,\tilde {y}_n ,\tilde {z}_n )$, $n =
1,...,N$. Let ${\rm {\bf r}}_{i,n} = {\rm {\bf \tilde {x}}}_n - {\rm
{\bf x}}_i $. Then the magnetic field at ${\rm {\bf \tilde {x}}}_n $
due to dipole ${\rm {\bf m}}_i $ is given by ${\rm {\bf H}}({\rm
{\bf m}}_i ,{\rm {\bf r}}_{i,n} )$, and the total magnetic field at
measurement point ${\rm {\bf \tilde {x}}}_n $ induced by all the $k$
dipoles is given by
\begin{equation}
{\bf H}_n ={\rm {\bf i}}{H}_{n,x} + {\rm {\bf j}} {H}_{n,y} + {\rm
{\bf k}} {H}_{n,z} = {\bf i} \sum_{i=1}^k  H_x({\bf m}_i, {\bf
r}_{i,n} ) + {\bf j} \sum_{i=1}^k  H_y({\bf m}_i, {\bf r}_{i,n} )+
{\bf k} \sum_{i=1}^k H_z({\bf m}_i, {\bf r}_{i,n} )
\end{equation}
The observed magnetic field at ${\rm {\bf \tilde {x}}}_n $ is ${\rm
{\bf \tilde {H}}}({\rm {\bf \tilde {x}}}_n ) \equiv {\rm {\bf \tilde
{H}}}_n = {\rm {\bf i}}\tilde {H}_{n,x} + {\rm {\bf j}}\tilde
{H}_{n,y} + {\rm {\bf k}}\tilde {H}_{n,z} $. Assuming an independent
Gaussian noise with standard deviation $\sigma $ in each of the
measured components, the likelihood function is then
\begin{equation}
\pi ({\bf \tilde H} | {\rm {\bf \Theta }}_k ) \propto
\frac{1}{\sigma^{3n}} exp \left (  -\frac{\sum_{n=1}^N \left (
(H_{n,x}-\tilde H_{n,x})^2  +(H_{n,y}-\tilde H_{n,y})^2
 +(H_{n,z}-\tilde H_{n,z})^2 \right)}{2\sigma^2}
 \right )
\end{equation}
\noindent where ${\rm {\bf \Theta }}_k = (\omega _1 ,\varphi _1 ,m_1
,x_1 ,y_1 ,z_1 ,...,\omega _k ,\varphi _k ,m_k ,x_k ,y_k ,z_k )$
denotes the model, with each dipole ${\rm {\bf m}}_i $ having six
parameters $(\omega _i ,\varphi _i ,m_i ,x_i ,y_i ,z_i )$
representing its direction, strength and location. The first two
parameters $(\omega _i ,\varphi _i )$ are the spherical polar
coordinates of the unit vector for the dipole, i.e. $\hat {m}_{i,x}
= \sin \omega _i \cos \varphi _i $, $\hat {m}_{i,y} = \sin \omega_i
\sin \varphi_i $, $\hat {m}_{i,z} = \cos \omega _i $\textbf{.}

In a geophysical context, the object creating the magnetic anomaly
could be represented by a collection of dipoles with the same
orientation ${\rm {\bf \hat {m}}} = (\omega ,\varphi )$ and the same
strength $m$. In such a case the parameter vector for a collection
of $k$ dipoles is ${\rm {\bf \Theta }}_k = (\omega ,\varphi ,m,x_1
,y_1 ,z_1 ,...,x_k ,y_k ,z_k )$, i.e. there are only $3k + 3$
parameters for the model of $k$ dipoles.

\begin{figure}[htbp]
\centerline{\includegraphics[width=7cm,
height=5.5cm]{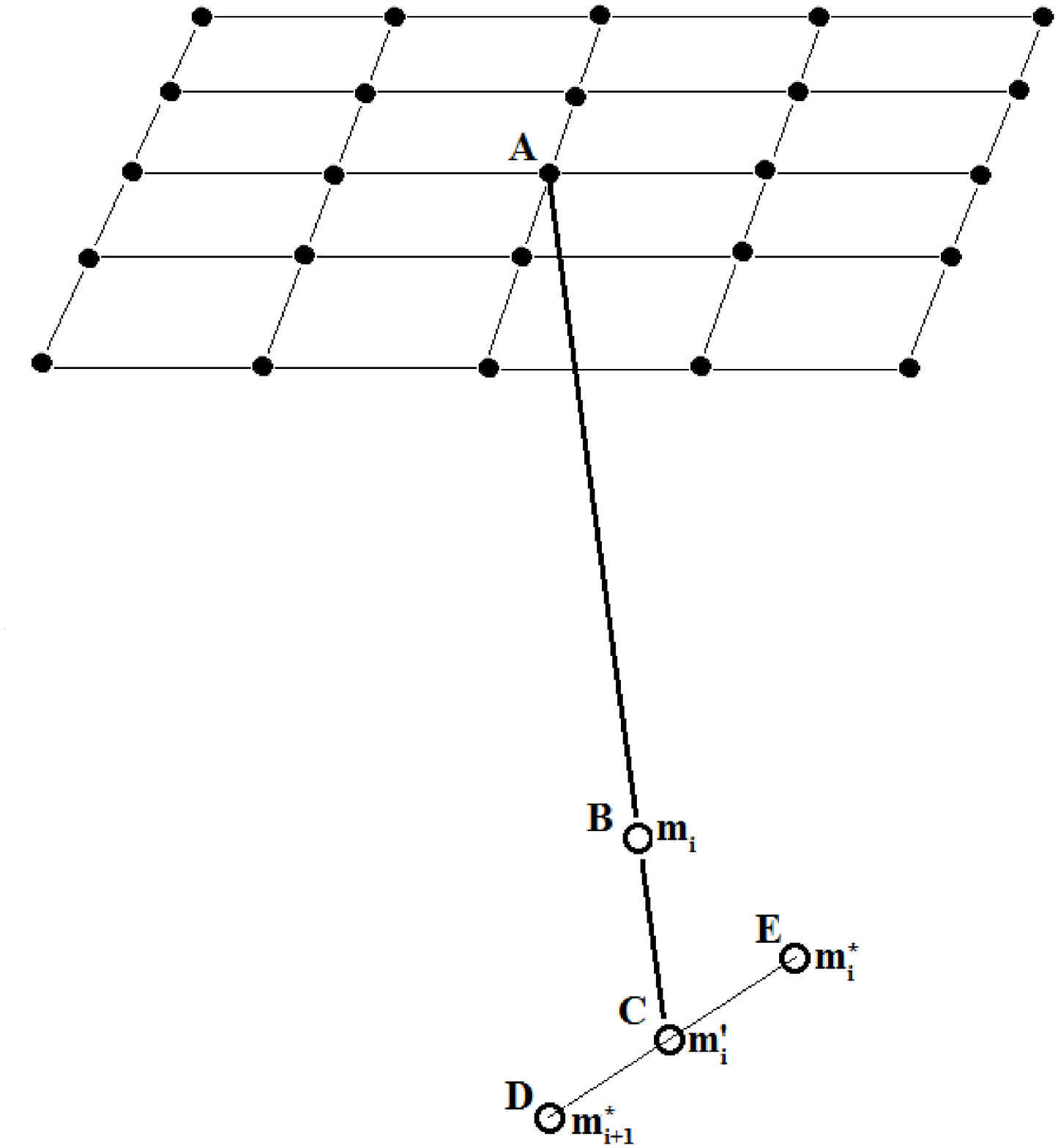}}
\caption{Illustration of dipoles in a birth-move.} \label{fig1}
\end{figure}

\section{REVERSIBLE JUMP MCMC ALGORITHM}

We now describe a reversible jump MCMC algorithm for the dipole
model. First, we describe the within-model moves where the number of
dipoles is fixed at $k$, i.e. there are no birth nor death moves.

\subsection{Within model moves - Metropolis-Hastings algorithm}

The Metropolis-Hastings algorithm was first described by
\citet{Hastings1970} as a generalization of the Metropolis
algorithm,  \citet{MetropolisEta1953}. Denote the state vector for
the model of $k$ dipoles as
\[
{\rm {\boldsymbol{\theta} }} = (\theta _1 ,\theta _2 ,...,\theta
_{3k + 2} ,\theta _{3k + 3} ) = {\rm {\bf \Theta }}_k = (\omega
,\varphi ,m,x_1 ,y_1 ,z_1 ,...,x_k ,y_k ,z_k ).
\]
At step $t$ the state vector ${\boldsymbol  \theta } = {\boldsymbol
\theta }^{(t)}$ and we wish to update it to a new state ${\rm {\bf
\theta }}^{(t + 1)}$. We generate a candidate ${\boldsymbol \theta
}^* $ from candidate generating density $q({\boldsymbol \theta }^ *
\vert {\boldsymbol \theta }^{(t)})$, we then accept this point as
the new state of the chain with probability $p_a ({\boldsymbol
\theta }^{(t)},{\boldsymbol \theta }^ * )$ given by
\begin{equation}
p_a({\boldsymbol  \theta } ^{(t)},{\boldsymbol \theta}^*) =
\text{min} \left \{1, \frac{\pi({\bf \tilde H} | {\boldsymbol
\theta}^* )\pi({\boldsymbol \theta}^* )q({\boldsymbol \theta^{(t)} }
| {\boldsymbol \theta}^* )}{\pi({\bf \tilde H} | {\boldsymbol
\theta^{(t)}} )\pi({\boldsymbol \theta^{(t)}} )q({\boldsymbol \theta
}^*  | {\boldsymbol \theta^{(t)}} )} \right \}
\end{equation}
\noindent where $\pi ({\rm {\bf \tilde {H}}}\vert {\boldsymbol
\theta })$ is the likelihood given by (3), $\pi ({\boldsymbol \theta
})$ is the prior density. If the proposal is accepted, we let the
new state ${\boldsymbol \theta }^{(t + 1)} = {\boldsymbol  \theta }^
* $, otherwise ${\boldsymbol \theta }^{(t + 1)} = {\boldsymbol \theta
}^{(t)}$. It is often more efficient to partition the state variable
${\boldsymbol \theta }$ into components and update these components
one by one. This was the framework for MCMC originally proposed by
Metropolis et al. (1953), and it is used in this work. For each
component $\theta _j $, we take the normal density as the proposal
density $q(\theta _j^\ast \vert \theta _j^{(t)} ) = f_n (\theta
_j^\ast - \theta _j^{(t)} \vert 0,\sigma _j )$, where $f_n (.\vert
0,\sigma _j )$ is the normal density with zero mean and standard
deviation $\sigma _j $. A sensible choice for the $\sigma _j $
values is to let $\sigma _1 = \sigma _2 = \sigma _\omega $ for the
two common polar coordinates, $\sigma _3 = \sigma _m $ for the
common magnetic strength, and $\sigma _4 = \sigma _5 = ... = \sigma
_{3k + 3} = \sigma _{xyz} $ for all the position coordinates.

\subsection{Trans-dimensional moves}

The reversible jump Markov chain Monte Carlo  proposed by
\citet{Green1995} provides a framework for constructing reversible
Markov chain samplers that jump between parameter spaces of
different dimensions, thus permitting exploration of joint parameter
and model probability space via a single Markov chain. As shown by
\citet{Green1995}, detailed balance is satisfied if the proposed
move from ${\rm {\bf \Theta }}_i $ to ${\rm {\bf \Theta }}_j $ is
accepted with probability $\alpha = \min \left\{ {1,\alpha _{i \to
j} ({\rm {\bf \Theta }}_i ,{\rm {\bf \Theta }}_j )} \right\}$, with
$\alpha _{i \to j} ({\rm {\bf \Theta }}_i ,{\rm {\bf \Theta }}_j )$
given by
\begin{equation}
\alpha_{i \to j}({\rm {\bf \Theta }}_i ,{\rm {\bf \Theta }}_j ) =
\frac{\pi({\bf \tilde H} | {\rm {\bf \Theta }}_j )r_{j \to i} ({\rm
{\bf \Theta }}_j )\varphi_j({\rm {\bf {u}}_j }  | {\rm {\bf {\beta
}}_j} )}{\pi({\bf \tilde H} | {\rm {\bf \Theta }}_i ) r_{i \to j}
({\rm {\bf \Theta }}_i )\varphi_i({\rm {\bf {u}}_i }  | {\rm {\bf
{\beta }}_i} )} \left|\frac{\partial g_{i \to j} ({\rm {\bf \Theta
}}_i, {\bf u}_i )}{\partial ({\rm {\bf \Theta }}_i, {\bf u}_i )}
\right |
\end{equation}
\noindent where $r_{i \to j} ({\rm {\bf \Theta }}_i )$ is the
probability that a proposed jump from ${\rm {\bf \Theta }}_i $ to
${\rm {\bf \Theta }}_j $ is attempted, $\varphi _i (.)$ is a
proposal density, and $\left| {\partial g_{i \to j} ({\rm {\bf
\Theta }}_i ,{\rm {\bf u}}_i ) / \partial ({\rm {\bf \Theta }}_i
,{\rm {\bf u}}_i )} \right|$ is the Jacobian of the deterministic
mapping. Efficiency of RJ-MCMC depends on the choice of mapping
function $g_{i \to j} $ and the proposal density $\varphi _i (.)$.

\subsubsection{Birth move}

Typically, a birth-move is from ${\rm {\bf \Theta }}_k $ to ${\rm
{\bf \Theta }}_{k + 1} $, i.e. in the above description we have $i =
k$ and $j = k + 1$. In order to have a reasonable acceptance rate
for the birth move, we try to keep the change in the likelihood
function from $\pi ({\rm {\bf \tilde {H}}}\vert {\rm {\bf \Theta
}}_k )$ to $\pi ({\rm {\bf \tilde {H}}}\vert {\rm {\bf \Theta }}_{k
+ 1} )$ to be small, i.e. the birth-move is designed in such a way
that $\pi ({\rm {\bf \tilde {H}}}\vert {\rm {\bf \Theta }}_k )
\approx \pi ({\rm {\bf \tilde {H}}}\vert {\rm {\bf \Theta }}_{k + 1}
)$. To achieve this small perturbation in likelihood function,
instead of randomly adding a new dipole to the system, we replace
one of the existing dipoles with two new dipoles. Ideally, the
combined magnetic field produced by the two new dipoles should be
very close to the magnetic field of the replaced dipole, at every
measurement point ${\rm {\bf \tilde {x}}}_n $, $n = 1,...,N$. It is
analytically difficult to ensure this closeness of magnetic field at
every measurement point ${\rm {\bf \tilde {x}}}_n $.

We can simplify the problem by ensure that the magnetic field
produced by the new pair of dipoles is close to that of the old
dipole at one key measurement point ${\rm {\bf \tilde {x}}}_a $, $1
\le a \le N$. Typically the measurement points can be arranged in a
horizontal ($\tilde {z} = \mbox{const})$ rectangular lattice$\tilde
{x}_{\min } \le \tilde {x} \le \tilde {x}_{\max } $, $\tilde
{y}_{\min } \le \tilde {y} \le \tilde {y}_{\max } $, as shown in
Figure 1, and ${\rm {\bf \tilde {x}}}_a $ can be chosen to be
located at the centre of the lattice.

In figure 1, the key measurement point ${\rm {\bf \tilde {x}}}_a $
is marked as A. Assuming the randomly chosen dipole ${\rm {\bf m}}_i
$ is located at point B with coordinate vector ${\rm {\bf x}}_i $,
we wish to find two locations near B such that the new pair of
dipoles located at these two points will produce a combined magnetic
field close to that of the old dipole ${\rm {\bf m}}_i $. Let the
two new locations be E and D for the new dipoles ${\rm {\bf
m}}_i^\ast $ and ${\rm {\bf m}}_{i + 1}^\ast $, as shown in Figure
1.

Denote the vector $\overrightarrow {AB} = - {\rm {\bf r}}_{B,A}= -
r_{B,A} {\rm {\bf \hat r}}_{B,A}$, where $r_{B,A}$ is the distance
between A and B and ${\rm {\bf \hat {r}}}_{B,A}$ is the unit vector
from B to A (from dipole to measurement point). Now extend
$\overrightarrow {AB} $ to $\overrightarrow {AC} $ such that
$\overrightarrow {AC} = - {\rm {\bf r}}_{C,A} = - \sqrt[3]{2}\times
r_{B,A} {\rm {\bf \hat r}}_{B,A}$, i.e. let C be on the same line as
$\overrightarrow {AB} $ and the length of $\overrightarrow {AC} $ is
$\sqrt[3]{2}$ times that of the length of $\overrightarrow {AB} $.
Now we put two dipoles $({\rm {\bf {m}'}}_i ,{\rm {\bf {m}'}}_i )$
at the same location C.

We now can easily show that a pair of dipoles $({\rm {\bf {m}'}}_i
,{\rm {\bf {m}'}}_i )$ co-located at C produce a combined magnetic
field (all 3 components) at measurement point A identical to that of
dipole ${\rm {\bf m}}_i $, given that all dipoles have the same
strength $m$ and unit vector ${\rm {\bf \hat {m}}}$. Applying field
equation (3) to dipole ${\rm {\bf m}}_i $ and measurement location
$A$, we have
\begin{equation}
{\bf H}({\bf m}_i, {\bf r}_{B,A}) =\frac{\mu_0m}{4\pi r^3_{B,A}}
\left ( 3({\bf \hat m} \cdot {\bf \hat r}_{B,A})({\bf r}_{B,A} -
{\bf \hat m} )\right )
\end{equation}
Similarly, applying field equation (3) to dipole ${\rm {\bf
{m}'}}_i$  and measurement location A, we have
\begin{equation}
{\bf H}({\bf m}_i', {\bf r}_{C,A}) =\frac{\mu_0m}{4\pi r^3_{C,A}}
\left ( 3({\bf \hat m} \cdot {\bf \hat r}_{C,A})({\bf r}_{C,A} -
{\bf \hat m} )\right )
\end{equation}
Because $r_{C,A}= \sqrt[3]{2}\times r_{B,A}$ and ${\rm {\bf \hat
{r}}}_{C,A} = {\rm {\bf \hat {r}}}_{B,A}$, comparing (6) and (7) we
obtain ${\rm {\bf H}}({\rm {\bf {m}'}}_i^ ,{\rm {\bf r}}_{C,A} ) =
{\rm {\bf H}}({\rm {\bf m}}_i ,{\rm {\bf r}}_{C,A} ) / 2$. Thus a
pair of dipole $({\rm {\bf {m}'}}_i ,{\rm {\bf {m}'}}_i )$
co-located at point C produce combined magnetic field at point A
identical to the magnetic field produced by dipole ${\rm {\bf m}}_i
$ located at B, provided all the three dipoles have a common
strength $m$ and unit vector ${\rm {\bf \hat {m}}}$.

Therefore we propose the following birth-move procedure, assuming
the key measurement point A is fixed throughout the MCMC iterations

\begin{enumerate}

\item Randomly remove a dipole ${\bf m}_i$ (located at B in Figure
1) .
\item   Locate point C by extending the line from  $\overrightarrow {AB} $ to $\overrightarrow {AC}$,
so that C is on the same line as  $\overrightarrow {AB} $ and
$|\overline{AB}| = \sqrt[3]{2}|\overline{AC}|$;
\item   Put two dipoles at location C;
\item Generate three independent random variables  $dx$, $dy$, $dz$ from normal
distribution $f_n (0,\sigma _{xyz} )$;
\item  Move one of the two dipoles from C to location E, such that $\overrightarrow {CE} = {\rm {\bf i}}dx + {\rm {\bf j}}dy +
{\rm {\bf k}}dz$. This new dipole is denoted as ${\rm {\bf
m}}_i^\ast$ located at ${\rm {\bf x}}_i^\ast$ (point E in Figure 1);
\item  Move the other dipole from C to location D, such that
$\overrightarrow {CD} = - \overrightarrow {CE}$. This new dipole is
denoted as  ${\rm {\bf m}}_{i+1}^\ast$  located at ${\rm {\bf
x}}_{i+1}^\ast$  (point D in Figure 1).

\end{enumerate}

The random vector ${\rm {\bf u}}_k $ corresponding to the birth-move
from ${\rm {\bf \Theta }}_k $ to ${\rm {\bf \Theta }}_{k + 1} $ is
identified as ${\rm {\bf u}}_k = (dx,dy,dz)$ with a single parameter
$\beta _k = \sigma _{xyz} $, the standard deviation for the random
walk of a dipole. Thus
\[
\varphi _k ({\rm {\bf u}}_k \vert {\rm {\bf \beta }}_k ) = f_n
(dx\vert 0,\sigma _{xyz} )f_n (dy\vert 0,\sigma _{xyz} )f_n (dz\vert
0,\sigma _{xyz} ).
\]
To find the Jacobian of the deterministic mapping, we first find the
mapping function $g_{k \to k + 1} ({\rm {\bf \Theta }}_k ,{\rm {\bf
u}}_k )$
\begin{equation}
x_i^\ast=(1-c)\tilde{x}_a + cx_i +dx,\;\;\;
y_i^\ast=(1-c)\tilde{y}_a + cy_i +dy,\;\;\;
z_i^\ast=(1-c)\tilde{z}_a + cz_i +dz
\end{equation}
\begin{equation}
x_{i+1}^\ast=(1-c)\tilde{x}_a + cx_i -dx,\;\;\;
y_{i+1}^\ast=(1-c)\tilde{y}_a + cy_i -dy,\;\;\;
z_{i+1}^\ast=(1-c)\tilde{z}_a + cz_i -dz
\end{equation}
\noindent where $c = \sqrt[3]{2}$, from which we find the Jacobian
to be $\left| {\partial g_{k \to k + 1} ({\rm {\bf \Theta }}_k ,{\rm
{\bf u}}_k ) /
\partial ({\rm {\bf \Theta }}_k ,{\rm {\bf u}}_k )} \right| = 8c^3 = 16$.
Assume the probability to propose the general birth move (as against
a birth-move or a within-model move) is $p_b $, and we know the
probability of choosing ${\rm {\bf m}}_i $ among the $k$ dipoles is
$1 / k$, so for the birth-move we have $r_{k \to k + 1} ({\rm {\bf
\Theta }}_k ) = p_b / k$ and
\begin{equation}
\frac{1}{r_{k \to k + 1}({\rm {\bf \Theta }}_k)\varphi_k({\rm {\bf
{u}}_k }  | {\rm {\bf {\beta }}_k} )} \left| \frac{\partial g_{k \to
k + 1} ({\rm {\bf \Theta }}_k ,{\rm {\bf u}}_k ) }{
\partial ({\rm {\bf \Theta }}_k ,{\rm {\bf u}}_k ) } \right|
=\frac{16k}{p_bf_n (dx\vert 0,\sigma _{xyz} )f_n (dy\vert 0,\sigma
_{xyz} )f_n (dz\vert 0,\sigma _{xyz} )}
\end{equation}

\subsubsection{Death move}

This is the reversal of the birth-move (still using Figure 1 as
illustration):
\begin{enumerate}
\item Randomly select a pair of dipoles among the $k(k - 1) / 2$ pairs,
delete them from the system, assuming that, without losing
generality, the pair are ${\rm {\bf m}}_j ({\rm {\bf x}}_j )$
located at E and ${\rm {\bf m}}_{j + 1} ({\rm {\bf x}}_{j + 1} )$
located at D;
\item find the middle point C between E and D, as shown in
Figure 1, and locate point B on the line $\overline {AC} $ so that
$\left| {\overline {AC} } \right| = \sqrt[3]{2}\times \left|
{\overline {AB} } \right|$.
\item Put one dipole $m_j^\ast ({\rm {\bf x}}_j^\ast )$ at location B,
where ${\rm {\bf x}}_j^\ast $ is the coordinate of point B.
\end{enumerate}
In the above one-for-two death-move, the only random number is from
uniform $\left( 1, k(k - 1) / 2 \right)$. The probability of making
the specific death-move is $r_{k + 1} ({\rm {\bf \Theta }}_{k + 1} )
= 2p_d / \left( {k(k - 1)} \right)$, where $p_d $ is the probability
of attempting a general death-move. The mapping function $g_{k + 1
\to k} ({\rm {\bf \Theta }}_{k + 1} ,{\rm {\bf u}}_{k + 1} )$ is the
inverse of the mapping function $g_{k \to k + 1} ({\rm {\bf \Theta
}}_k ,{\rm {\bf u}}_k )$.

\subsubsection{Acceptance rates}

Combining birth-move and death-move as described above, we obtain
the following expressions for acceptance rates:

\textbf{\textit{Birth-move acceptance rate}}
\begin{equation}
p_a({\rm {\bf \Theta }}_k ,{\rm {\bf \Theta}}_{k+1}
)=\text{min}\{1,\alpha_{k \to k + 1}\}
\end{equation}
\begin{equation}
\alpha_{k \to k + 1}=\frac{32p_d\times \pi({\bf \tilde H} | {\rm
{\bf \Theta }}_{k+1} )} {p_b f_n (dx\vert 0,\sigma _{xyz} )f_n
(dy\vert 0,\sigma _{xyz} )f_n (dz\vert 0,\sigma _{xyz} ) (k-1)\times
\pi({\bf \tilde H} | {\rm {\bf \Theta }}_k )}
\end{equation}
\textbf{\textit{Death-move acceptance rate}}
\begin{equation}
p_a({\rm {\bf \Theta }}_{k+1} ,{\rm {\bf \Theta}}_k
)=\text{min}\{1,\alpha_{k+1 \to k}\}
\end{equation}
\begin{equation}
\alpha_{k+1 \to k}=\frac{p_b f_n (dx\vert 0,\sigma _{xyz} )f_n
(dy\vert 0,\sigma _{xyz} )f_n (dz\vert 0,\sigma _{xyz} ) (k-1)\times
\pi({\bf \tilde H} | {\rm {\bf \Theta }}_k )} {32p_d\times \pi({\bf
\tilde H} | {\rm {\bf \Theta }}_{k+1} )}
\end{equation}

\section{PRELIMINARY RESULTS}

We consider three cases: 1 - A bulky formation; 2 - A thin plate; 3
- Two objects. In each case we start with a single dipole, located
at an arbitrary depth below the measured magnetic field, and with an
arbitrary orientation and a fixed strength.
\begin{figure}[htbp]
\centerline{\includegraphics[width=2.4in,height=1.9in]{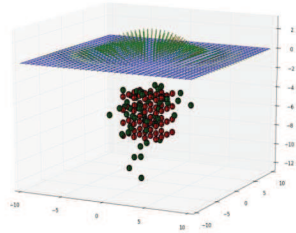}}
\caption{Predicted dipoles and their vector fields for case 1.}
\label{fig2}
\end{figure}

\textbf{Case 1.}
 In this case the dipoles form a regular cube.
Figure 2 shows a sample after 50000 simulations. In Figure 2, the
red balls represent the true model, and green balls are the 'best'
prediction. The horizontal blue lattice indicates measurement
points, and the lines originating from these points are the magnetic
vectors with green corresponding to the green dipoles (the predicted
dipoles) and red corresponding to the red diploes (the true model).
As can be seen, on the whole, the inversed dipoles reasonably
assemble the true model, with a few dipoles drifting to the deeper
depth. The predicted magnetic vector field matches that of the true
model very well - the green vectors and red vectors appear to be the
same everywhere on the measurement lattice.

\textbf{Case 2.} In this case, the dipoles form a horizontal thin
sheet, as shown by the blue balls in Figure 3. As seen in figure 3,
the resulting dipoles are too much scattered vertically.
Nevertheless, the horizontal scattering of the dipoles resemble the
true model, and the resulting magnetic field vector matches the true
vector field very well.

\textbf{Case 3.} In this case the dipoles form two separated
identical cubes at a significantly different depths and horizontal
locations. It can be seen there are still too many dipoles scattered
in between the two objects, and already the fit between the
predicted and measured magnetic vector fields are very good.
\begin{figure}[htbp]
\centerline{\includegraphics[width=2.4in,height=1.9in]{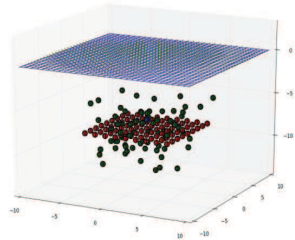}}
\caption{Predicted dipoles and their vector fields for case 2.}
\label{fig3}
\end{figure}
\begin{figure}[htbp]
\centerline{\includegraphics[width=2.4in,height=1.9in]{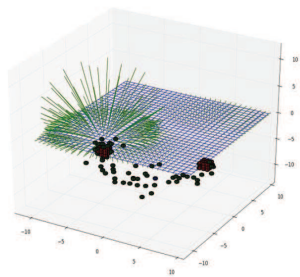}}
\caption{Predicted dipoles and their vector fields for case 3.}
\label{fig4}
\end{figure}

The three test cases show that the present method is promising but
some challenges remain. For a single cube-like object, the inverse
is not too bad in terms of representing the overall shape of the
object, although there seem to be always some dipoles scattered
below the object.  If the single object is a bit extreme, such as a
horizontal thin sheet, the inverse cannot predict the depth
resolution -- it is too scattered vertically.  For a more
challenging problem of two objects located at different depth, the
inverse tried hard to locate both, but with too many dipoles
scattered in between.

In all the above cases the forward problem was well resolved -- i.e.
the predicted magnetic field agrees very well with measured field.
This is typical in geophysical inversion -- non-uniqueness or
ill-conditioning is demonstrated in terms of large uncertainties in
the prediction of depth.

\section{CONCLUSIONS}
\label{sec:conclusionsomparison} We have proposed a reversible jump
Markov chain Monte Carlo (RJ-MCMC) algorithm for both the magnetic
vector and its gradient tensor to deal with this trans-dimensional
inverse problem where the number of unknowns is one the unknowns.  A
special birth-death move strategy is designed to obtain a reasonable
rate of acceptance for the RJ-MCMC sampling. Some preliminary
results show the strength and challenges of the algorithm in
inversing the magnetic measurement data. Although it is very
difficult, if not impossible, to predict each individual dipole
accurately, it is important to predict the cloud of dipoles
accurately (e.g. uniformly distributed with the edges close to the
true boundary). A different likelihood function or prior (in
addition to new method) may be of help in this regard. As always, it
is difficult to predict the depth of the object with reasonable
certainty. Better ways are needed to locate multiple objects - with
a clean break between two distinct objects, especially when they are
located at very different depths.

\section*{Acknowledgement}
This project was funded by the Capability Development Fund of CSIRO
Earth Science and Resource Engineering.

\end{document}